\documentclass[aps,prl,doublecolumn,superscriptaddress,floatfix]{revtex4-1}
\usepackage[utf8]{inputenc}
\usepackage[T1]{fontenc}
\usepackage{graphicx}%
\usepackage{amsmath}
\usepackage{amsfonts}
\usepackage{amssymb}

\begin{document}


\title{Spin-filter effect at the interface of magnetic/non-magnetic
  homojunctions in Li doped ZnO nanostructures}

\keywords{Spin filter, Defect Induced Magnetism, Giant
  Magnetoresistance, ZnO, Nanowire, Magnetic Homojunction}

\author{L. Botsch} \email{lukas.botsch@uni-leipzig.de}
\affiliation{Division of Superconductivity and Magnetism,
  Felix-Bloch-Institute for Solid State Physics, University of
  Leipzig, Linn\'estr. 5, D-04103 Leipzig, Germany}

\author{I. Lorite} \affiliation{Division of Superconductivity and
  Magnetism, Felix-Bloch-Institute for Solid State Physics, University
  of Leipzig, Linn\'estr. 5, D-04103 Leipzig, Germany}

\author{Y. Kumar} \altaffiliation[Present address: ]{Solid State
  Physics Division, Bhabha Atomic Research Centre, Mumbai 400085,
  India} \affiliation{Division of Superconductivity and Magnetism,
  Felix-Bloch-Institute for Solid State Physics, University of
  Leipzig, Linn\'estr. 5, D-04103 Leipzig, Germany}

\author{P. Esquinazi} \affiliation{Division of Superconductivity and
  Magnetism, Felix-Bloch-Institute for Solid State Physics, University
  of Leipzig, Linn\'estr. 5, D-04103 Leipzig, Germany}

\author{T. Michalsky} \affiliation{Division of Semiconductor Physics,
  Felix-Bloch-Institute for Solid State Physics, University of
  Leipzig, Linn\'estr. 5, D-04103 Leipzig, Germany}

\author{J. Zajadacz} \affiliation{Leibniz-Institut f\"ur
  Oberfl\"achenmodifizierung e.V., Permoserstr. 15, D-04318 Leipzig, Germany}

\author{K. Zimmer} \affiliation{Leibniz-Institut f\"ur
  Oberfl\"achenmodifizierung e.V., Permoserstr. 15, D-04318 Leipzig, Germany}


\begin{abstract}
  After more than a decade of extensive research on the magnetic order
  triggered by lattice defects in a wide range of nominally
  non-magnetic materials, we report its application in a spintronic
  device. This device is based on a spin-filter phenomenon we
  discovered at the interfaces between defect-induced magnetic and
  non-magnetic regions, produced at the surface of a Li doped ZnO
  microwire by low-energy proton implantation. Positive
  magnetoresistance is observed at 300~K and scales with the number of
  interfaces introduced along the wire.
\end{abstract}

\maketitle

\section{Introduction}
\label{sec:introduction}

Since the discovery of the giant magnetoresistance (GMR) effect in
magnetic/non-magnetic metal heterostructures nearly 30 years
ago~\cite{Baibich.1988,Binasch.1989}, researchers have focused on
finding new magnetoresistive effects and improving their strength for
applications in spintronic and sensor devices such as high-performance
read heads, nonvolatile memories, and other state-of-the-art storage
devices (see Ref.~[\onlinecite{Hartmann.2000}] and references therein).

In recent years, the advance in magnetic semiconductor material
research has opened new ways to achieve GMR in semiconductor
devices~\cite{Jin.2005, Xiong.2006, Rangaraju.2009}. Since the
prediction of ferromagnetism at room temperature in ZnO doped with
magnetic transitional metal ions~\cite{Dietl.2000}, it has become a
promising candidate material for semiconductor oxide based spintronic
applications, due to its large availability and low production cost.

Different approaches have been followed to induce GMR in ZnO based
material systems, for example transition metal (e.g. Co or Mn) doping
of ZnO thin films~\cite{Wang.2006, Tian.2008, Wang.2015},
magnetic/non-magnetic semiconductor oxide thin film
heterostructures~\cite{Tiwari.2006} and magnetic tunnel junctions
between a Co and a ZnO:Co film~\cite{Chen.2013}. All approaches showed
high magnetoresistance (MR), but only at low temperatures (below
$50\,\text{K}$).

We should note that after the first reports on magnetic order at room
temperature appeared, doubts were quickly raised on the homogeneity of
the ZnO-based samples as well as on the origin of the magnetism. In
the last years, however, the possibility to produce magnetic order by
introducing $\sim\,5\,\text{at\%}$ lattice defects (like Zn vacancies)
in ZnO has been succesfully shown in different
laboratories~\cite{Esquinazi.2013, Ogale.2010, Stoneham.2010,
  Schirmer.2006} and by different characterization methods including
X-ray magnetic circular dichroism (XMCD)~\cite{Lorite.2015}. This kind
of magnetism, called defect induced magnetism (DIM) has been
discovered in several other oxide and non-oxide
materials~\cite{Esquinazi.2013, Ogale.2010, Stoneham.2010,
  Schirmer.2006}. We also note that despite the large amount of
research on ZnO, no simple and feasible spintronic applications have
been found yet at room temperature.

In this work we present the observation of GMR in a ZnO based device
even at room temperature. The device consists of a Li-doped ZnO
microwire, in which DIM is induced by low energy proton
implantation~\cite{Lorite.2015}. Alternating magnetic and non-magnetic
regions are produced in a $\sim 10\,\text{nm}$ thick surface region of
the wire, effectively building magnetic/non-magnetic homojunctions
with sharp interfaces. We identify a spin filter mechanism at the
junction interfaces as the source of the magnetoresistance. In
contrast to other known magnetoresistive effects, the effect we
describe in this paper scales with the number of interfaces.

\section{Results and discussion}
\label{sec:results}

\begin{figure*}[ht]
  \centering
  \includegraphics[width=0.9\textwidth]{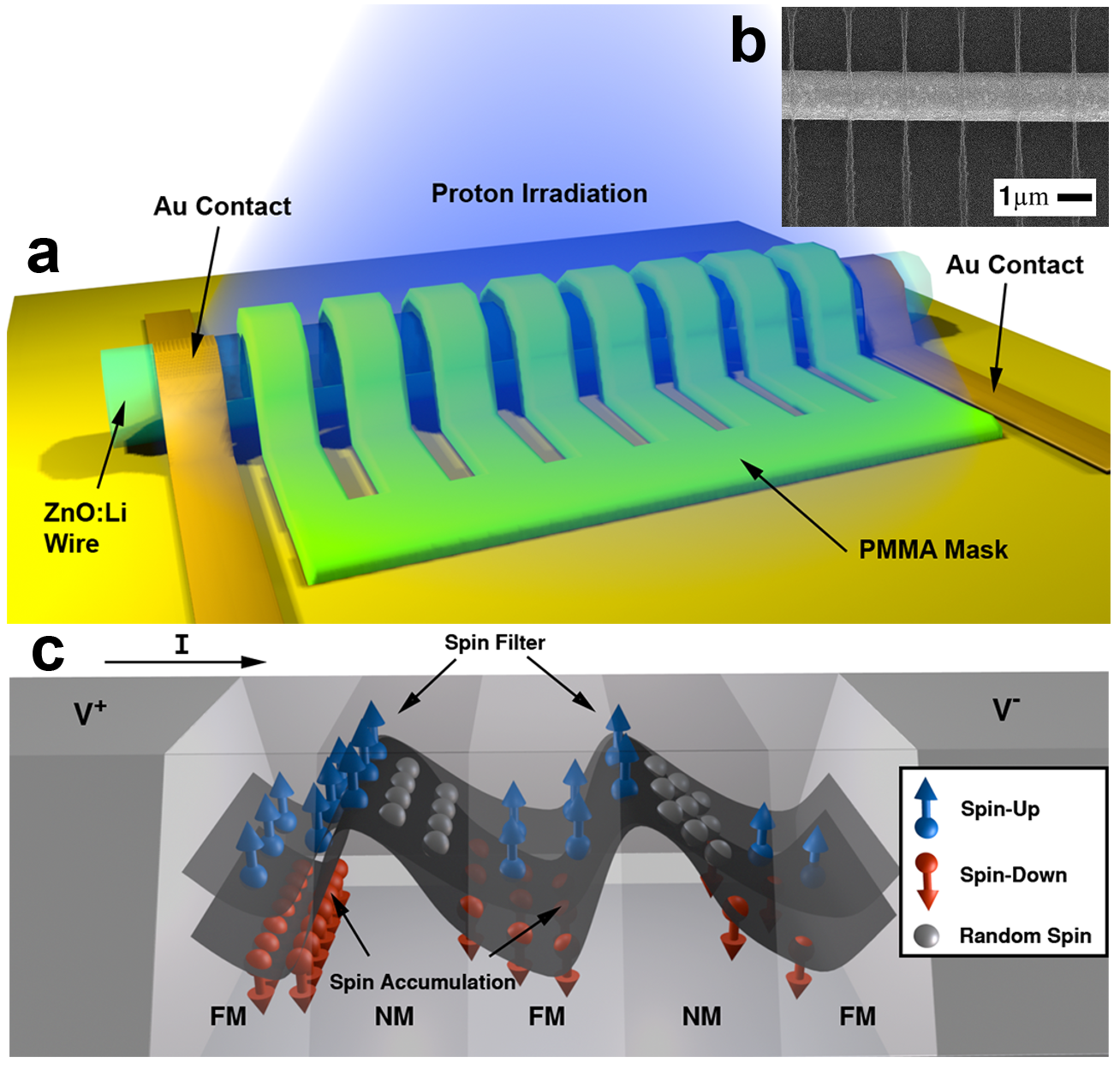}
  \caption{\label{fig:1} \textbf{Sample preparation and
      characterisation.}(a) Schematic of the $H^+$ implantation
    process to produce defect-induced magnetic strips on the surface
    of a ZnO:Li microwire. The PMMA mask is produced by electron beam
    lithography to achieve structure sizes down to 100~nm. (b) SEM
    picture of a wire with a mask covering 100~nm long regions for the
    creation of alternating magnetic/non-magnetic stripes by proton
    implantation. (c) Sketch describing the spin filtering effect at
    the interfaces between magnetically ordered (FM) and non-magnetic
    (NM) regions of the wire. In the FM regions, the current is spin
    polarized due to the spin subband splitting. Spin-down electrons
    accumulate at the FM/NM interface, while spin-up electrons can
    more easily pass the potential barrier. In the NM regions, spin-up
    and spin-down electrons become indiscernible and the current
    becomes unpolarized. At the NM/FM interface, the probability for
    spin-down electrons to be extracted from the NM region is higher
    than that of spin-up electrons. Thus, the current density
    decreases along the wire.}
\end{figure*}

Fig.~\ref{fig:1}~(a) depicts the process we used to create magnetic
strips at the surface of ZnO:Li microwires. The magnetic regions were
implanted with a proton dose of $10^{17}\,\text{cm}^{-2}$, while the
non-magnetic regions were implanted with a dose of
$10^{15}\,\text{cm}^{-2}$. For a detailed description of the process,
we refer the reader to the methods section. Photoluminescence scans
along the implanted wires clearly show the different regions (see
Supplementary Figure~1).

\begin{figure*}[ht]
  \centering
  \includegraphics[width=0.7\textwidth]{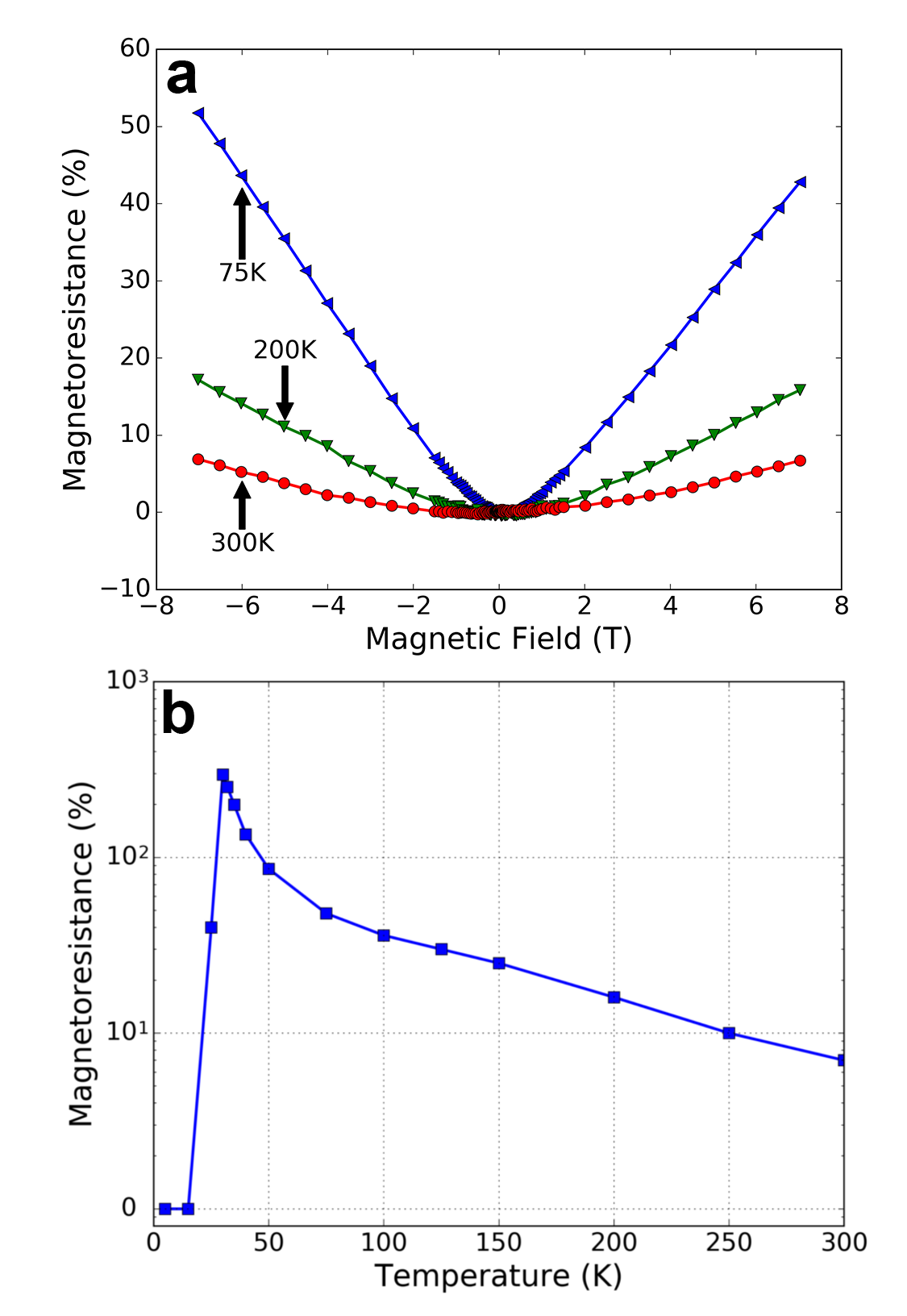}
  \caption{\label{fig:2} \textbf{Magnetoresistance effect.} (a)
    Magnetoresistance as a function of applied magnetic field of
    sample S1 with 37 magnetic strips measured at a temperature of
    $75\,\text{K}$, $200\,\text{K}$ and $300\,\text{K}$ while applying
    a bias voltage of $1\,\text{V}$. (b) MR as a function of
    temperature for the same wire at an applied magnetic field of
    $7\,\text{T}$ and bias voltage of $1\,\text{V}$.}
\end{figure*}

Fig.~\ref{fig:2}~(a) shows the field dependence of the MR at selected
temperatures between $75\,\text{K}$ and $300\,\text{K}$ and at an
applied bias voltage of $1\,\text{V}$ for the sample \textit{S1}. The
MR was calculated as follows:
\begin{equation}
\label{eq:1}
  \text{MR}(B,T) = \frac{R(B,T)-R(B=0,T)}{R(B=0,T)} \times 100,
\end{equation}
where $R(B=0,T)$ is the resistance at zero field and $R(B,T)$ the
resistance at a finite magnetic field $B$.

Fig.~\ref{fig:2}~(b) shows the MR as a function of temperature for
temperatures between $5$ and $300\,\text{K}$ when a bias voltage of
$1\,\text{V}$ and a magnetic field of $7\,\text{T}$ are applied. The
sample shows MR up to $300\,\%$ at $30\,\text{K}$ and $7\,\%$ at room
temperature. Above $50\,\text{K}$, the MR decreases exponentially.
Below $30\,\text{K}$, the MR drops within $\sim 10\,\text{K}$ to zero.
This behaviour is completely unusual. In general, the MR increases
while decreasing temperature. We will show in our simulations that
this behaviour can be explained by a ferromagnetic exchange
interaction.

As previously reported, the formation of Zn vacancies
($\text{V}_{\text{Zn}}$) during proton implantation is responsible for
the magnetic order in ZnO:Li microwires~\cite{Lorite.2015}. The
formation of such magnetic order is related to the density of
$\text{V}_{\text{Zn}}$ and their stabilization by Li dopants. A
certain concentration of $\text{V}_{\text{Zn}}$ is necessary to induce
magnetic order. Our previous experiments show that a dose of
$10^{17}\,\text{cm}^{-2}$ implanted protons induces a magnetic order,
while doses of $10^{15}-10^{16}\,\text{cm}^{-2}$ are not sufficient.
Thus, after implanting a dose of $10^{17}\,\text{cm}^{-2}$ protons in
the small regions along the wire, magnetic/non-magnetic (FM/NM)
interfaces are produced along the wire's main axis. Due to the short
penetration depth of the $\text{H}^+$, the interfaces between magnetic
and non-magnetic regions are sharply defined in space. In addition,
$\text{H}^+$ implantation introduces shallow donor
states~\cite{Bjrheim.2012}, which once activated provide free
electrons and decrease the resistivity of the wires in the implanted
region and within the first $10\,\text{nm}$ from the surface.

\begin{figure}[ht]
  \centering
  \includegraphics[width=0.7\textwidth]{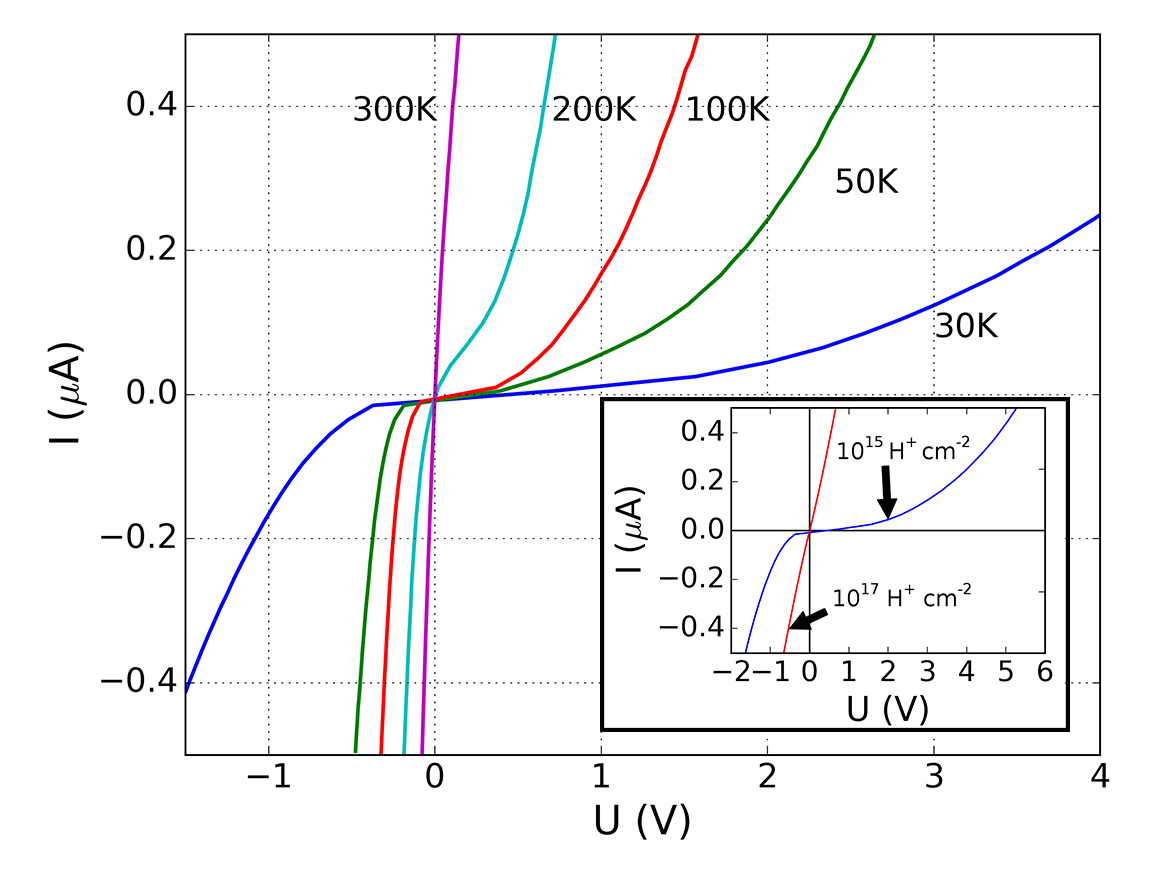}
  \caption{\label{fig:3} \textbf{IV characteristics} of the wire S1
    prepared with 37 magnetic stripes, measured without applied
    magnetic field in a temperature range between $30\,\text{K}$ and
    $300\,\text{K}$. The IV curves are non-linear and asymmetric due
    to the $\text{n}/\text{n}^+$ junctions formed by the weakly doped
    NM regions and the highly doped FM regions. The inset compares the
    IV characteristic at $30\,\text{K}$ of the wire \textit{S1} with
    37 magnetic strips, before (blue) and after (red) a uniform
    implantation of $10^{17}\,\text{cm}^{-2}$ $\text{H}^+$ along the
    whole wire. This last IV curve is linear, indicating that the
    nonlinearity indeed comes from the interfaces and not from the
    contacts.}
\end{figure}

Fig.~\ref{fig:3} shows the IV characteristics of sample \textit{S1}.
The IV curves show a clear diode rectification behavior. When the wire
is uniformly implantated with a dose of $10^{17}\,\text{cm}^{-2}$ of
$\text{H}^+$ along the whole wire, the IV curves become linear (see
inset of Fig.~\ref{fig:3}) at all temperatures. This is an indication that
the contacts are Ohmic. Thus, the non linearity observed is related to
a potential barrier produced between the highly donor doped
($\text{n}^+$) magnetic and the less donor doped (n) non-magnetic
regions, due to the very different free electron concentrations. Note
that the IV curves are asymmetric with respect to the current
polarity. In the present configuration, one would expect a symmetric
behaviour. This could be explained by an asymmetric termination of the
alternating chain of magnetic/non-magnetic regions at the metal
contacts.

As the IV characteristics of the sample are highly non-linear, we
measured the MR at different applied bias voltages (see supplementary
figure 2). Above 30~K, the MR is highest at small bias voltage and
decreases with increasing bias voltage. Below 30~K, the behaviour is
reversed. By applying a bias voltage, the potential barriers at the
interfaces between the FM and NM regions are shifted, indicating that
these interfaces play a role in the MR effect.

\begin{figure}[ht]
  \centering
  \includegraphics[width=0.7\textwidth]{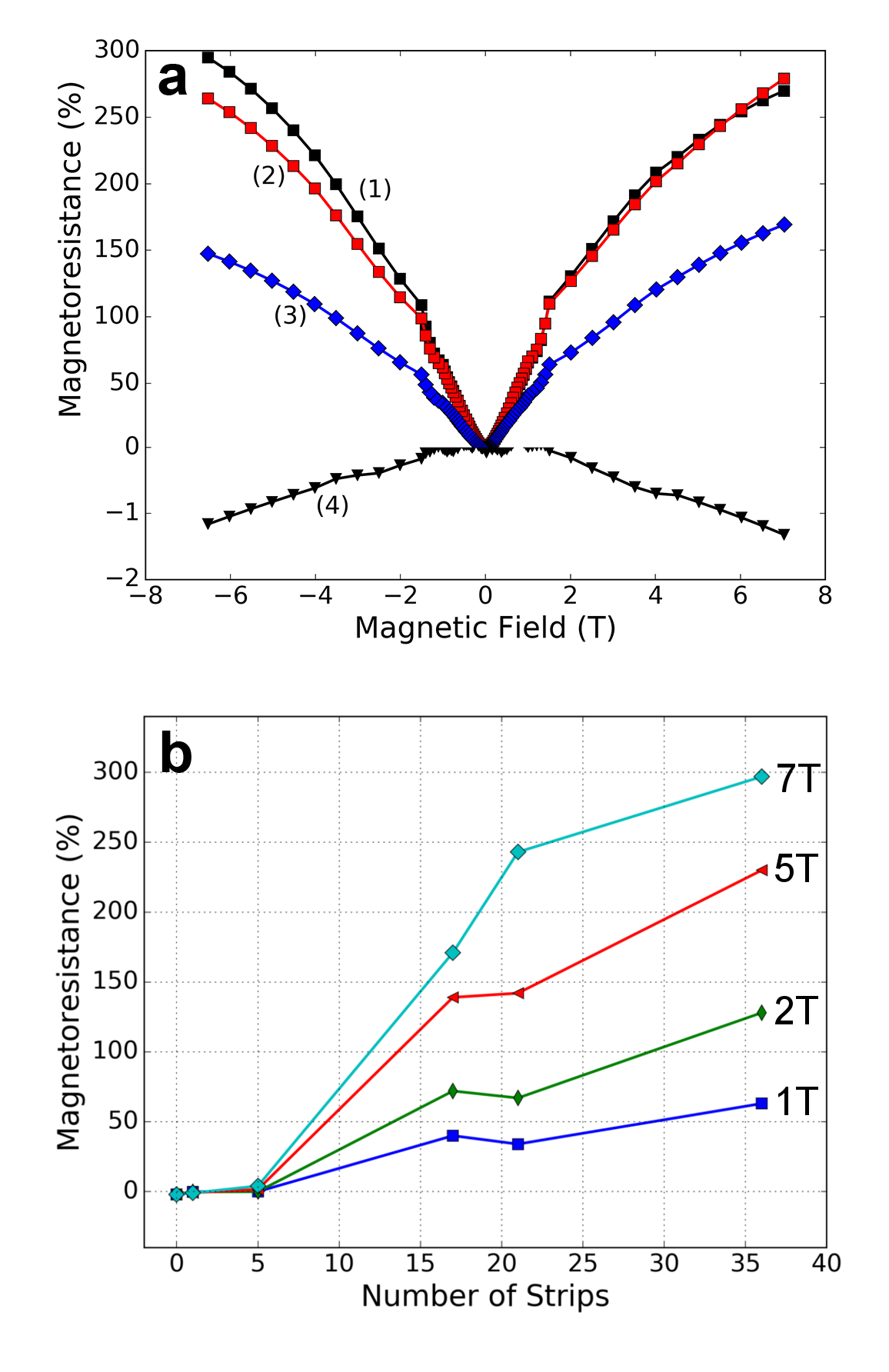}
  \caption{\label{fig:4} \textbf{Interface effect.} (a)
    Magnetoresistance measured at $30\,\text{K}$ with a bias of
    $1\,\text{V}$ as a function of applied magnetic field after an
    $\text{H}^+$ implantation dose of
    $2.5\times10^{15}\,\text{cm}^{-2}$ (1),
    $3.7\times10^{15}\,\text{cm}^{-2}$ (2),
    $7.5\times10^{15}\,\text{cm}^{-2}$ (3) and
    $3.7\times10^{16}\,\text{cm}^{-2}$ (4) in the non-magnetic regions
    of sample \textit{S1}. With increasing dose, the potential barrier
    height between the magnetic and non-magnetic regions becomes
    smaller and finally disappears after implanting a dose of
    $3.7\times10^{16}\,\text{cm}^{-2}$ (4), where the only remaining
    contribution to the MR comes from the magnetic regions. (b) MR at
    an applied field of $1\,\text{T}$, $2\,\text{T}$, $5\,\text{T}$
    and $7\,\text{T}$ as a function of the number of magnetic stripes.
    The results indicate that the effect scales with the number of
    interfaces. }
\end{figure}

To further prove the effect of the interfaces, we increased the
carrier concentration in the non-magnetic region of sample
\textit{S1}. For this purpose we implanted $\text{H}^+$ in the whole
wire. Note that the magnetic regions are already saturated with
$\text{H}^+$ and $\text{V}_{\text{Zn}}$ (from the previous
implantation step) so that the additional implantation process does
not change the conductivity of these regions significantly.
Fig.~\ref{fig:4}~(a) shows the magnetic field dependence of the MR for
different implantation doses in the NM regions. The curves (1), (2),
(3), (4) were measured after implanting doses of
$2.5\times10^{15}\,\text{cm}^{-2}$, $3.7\times10^{15}\,\text{cm}^{-2}$,
$7.5\times10^{15}\,\text{cm}^{-2}$ and
$3.7\times10^{16}\,\text{cm}^{-2}$ respectively. As one would expect,
the MR decreases with increasing implantation duration in the NM
regions, as the carrier concentration levels up with that in the FM
regions and the potential barriers shrink. After the implantation of
$3.7 \times 10^{16}\,\text{cm}^{-2}$ of protons in the NM regions, the MR
becomes negative (4) and has a maximum value of $-1\,\%$ at 7 T. The
potential barriers disappeared and with them the GMR effect. Note that
the observed negative MR is only due to the FM regions, as already
shown in a previous work~\cite{Lorite.2015c}.

This suggests that there are two different contributions, namely the
negative MR contribution from the magnetic regions and a giant
positive MR due to interfaces, in agreement with the variation of the
MR with the applied bias voltage. In such a case a decrease of the
number of interfaces along the wire would produce a reduction of the
total MR. To demonstrate it, we performed similar experiments on Li
doped ZnO wires on which different number of strips were produced.
Fig.~\ref{fig:4}~(b) shows the MR as a function of an applied magnetic
field at a constant temperature of $30\,\text{K}$ for different
samples having 36, 21, 17, 5, 1 and 0 magnetic strips. We can see that
the MR increases with the number of interfaces. For 1 strip, the
negative MR is observed, since the effect of a single interface is not
strong enough to compensate the contribution of the magnetic region.

The presented measurements show several peculiar features of the
effect at hand, namely: 1) The positive MR is linear in applied
magnetic field. 2) The positive MR scales with the number of NM/FM
interfaces. 3) The MR drops to zero at temperatures below
$30\,\text{K}$. These observations suggest that the measured MR
originates from a spin filter effect at the interfaces between the NM
and FM regions, as we will show.

Fig.~\ref{fig:1}~(c) shows a sketch of the model we use to explain the
observed effect. We first recognize that: 1) A potential barrier
exists between the magnetic and non-magnetic regions due to the
different carrier concentrations produced by the $\text{H}^+$
implantation. 2) In the magnetic (FM) strips, the spin degeneracy is
lifted and the conduction band is split due to the exchange interaction
($E_{ex}$) and a large Zeeman splitting ($E_Z$). All this induces a
spin polarization of the conduction electrons.

At each interface between the magnetic and non-magnetic regions, a
potential barrier is built due to the different doping profiles, which
results in a difference in chemical potential at both sides of the
interface. The spins can cross the potential barrier by thermionic
emission and undergo spin scattering processes in the non-magnetic
part. As the spin diffusion length in ZnO is smaller than the length
of the non-magnetic parts we created, the current remains less spin
polarized when reaching the next NM/FM interface.

At the NM/FM interface, a spin filtering process takes place, as one
spin orientation is energetically more favorable to cross the
interface, while the electrons with the opposite spin orientation get
accumulated at the interface due to the spin band splitting in the FM
strip. By adding more interfaces in series, the effect can potentially
be scaled indefinitely.

\begin{figure}[ht]
  \centering
  \includegraphics[width=0.7\textwidth]{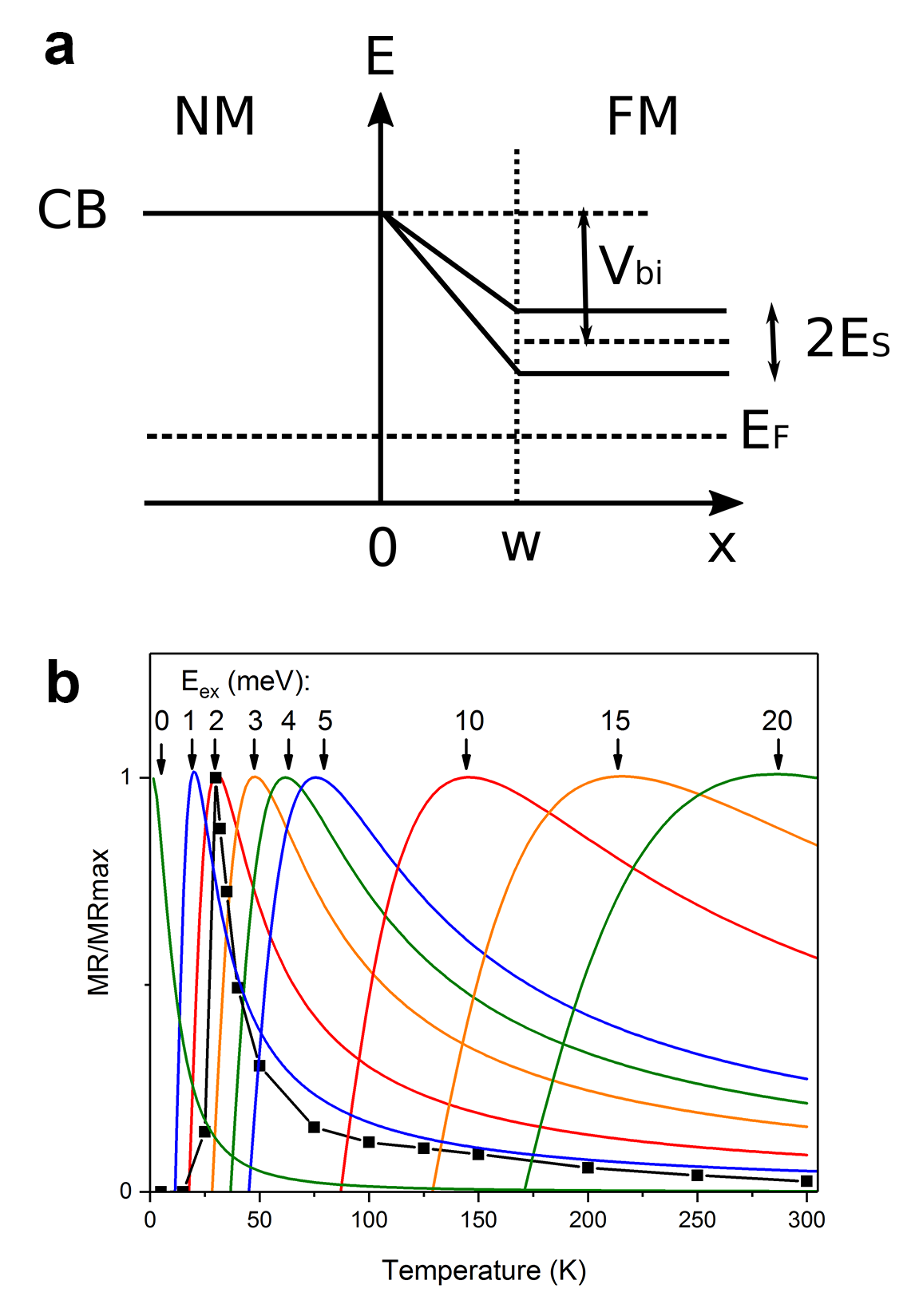}
  \caption{\label{fig:5} \textbf{Model of the spin filter effect.} (a)
    Sketch of the conduction band (CB) structure along the magnetic
    (FM) / non-magnetic (NM) junction. $V_{bi}$ is the builtin
    potential step, $E_F$ the Fermi level and $E_s$ the spin-band
    splitting energy. (b) The squares show the magnetoresistance at an
    applied magnetic field of 7~T and a bias voltage of 1~V measured
    as a function of temperature and normalized to its maximum value.
    The lines show the results of the numerical calculation of the
    temperature dependence of the MR due to a spin polarization
    (Eq.~(\ref{eq:thermionic-mr-model})) for values of the exchange
    energy $E_{ex}$ between 0~meV and 20~meV.}
\end{figure}

We can model the spin transport through these interfaces using a very
simple approximation of the barrier, as sketched in
Fig.~\ref{fig:5}~(a). At thermal equilibrium, the conduction band (CB)
is deformed in the magnetic region (FM) due to the different doping
concentrations $n_0$ and $N_0$ in the NM and FM regions respectively
and is shifted by
$V_{bi} = \frac{k_BT}{e}\log \left(\frac{N_0}{n_0}\right)$ compared to
the NM region. In the FM region, the spin-subbands are split by the
energy $2E_S = 2(E_{ex}+g\mu_BB)$, which has a magnetic field
dependent contribution from Zeeman interaction and a constant
contribution $E_{ex}$, the exchange energy. At the FM/NM interface, a 
space-charge region of width w is built out, in which spin-up and
spin-down electrons have different transmission probabilities. The
spin dependent current density through the interface can be expressed
as
\begin{equation}
  j_{\uparrow\downarrow} \propto \mu_{\uparrow\downarrow} T^2
  \exp\left[-\frac{e}{k_BT}\left(V_{bi}-V_n \pm \frac{E_s}{e}\right)\right]\left[\exp\left(\frac{eV_a}{k_BT}\right)-1\right]
\end{equation}
where $\mu_{\uparrow\downarrow}$ is the mobility of spin-up and -down
electrons respectively, $V_n$ the depth of the donor states taken from
the conduction band edge, $V_a$ the applied voltage, $e$ the electron
charge, $k_B$ the Boltzmann constant and $T$ the temperature. The
total charge current through the interface is then the sum of the two
spin channel currents $j = j_\uparrow + j_\downarrow$. From this, we
calculate the magnetoresistance:
\begin{equation}
\label{eq:thermionic-mr-model}
MR(B,T) = \frac{\tau_B(B,T)\left[\mu_\downarrow +
    \mu_\uparrow\tau_{ex}^4\right]\left[1+\left(\tau_B(B,T)\tau_{ex}(T)\right)^2\right]}{\left(1+\tau_{ex}(T)^2\right)\left[\mu_\downarrow
    + \mu_\uparrow\left(\tau_B(B,T)\tau_{ex}(T)\right)^4\right]} - 1
\end{equation}
with $\tau_B = \exp\left(\frac{g\mu_BB}{k_BT}\right)$ and
$\tau_{ex} = \exp\left(\frac{E_{ex}}{k_BT}\right)$ the Boltzmann
factors of the Zeeman and Exchange splitting respectively. In this
simple model, spin dependent scattering processes are encoded in the
electron mobility $\mu_{\uparrow\downarrow}$, that is assumed to
be different for the two spin orientations.

Fig.~\ref{fig:5}~(b) shows the normalized MR as a function of
temperature as calculated using Eq.~(\ref{eq:thermionic-mr-model}),
where we set the applied magnetic field to 7~T, the ratio of
mobilities of spin-up and -down electrons to
$\frac{\mu_\downarrow}{\mu_\uparrow} = 100$ and the exchange energy
$E_{ex}$ ranging between 0~meV and 20~meV. By comparing the
calculations to our measurements, we can estimate the exchange energy
of the magnetic regions of our sample to $E_{ex}\sim 1\,\text{meV}$.
Taking into account the Curie temperature reported for defect induced
magnetic ZnO (between 450~K and 700~K), this value for $E_{ex}$ fits
very well with those reported for several transition metals, e.g. Fe
($T_C=1041$~K, $E_{ex}=7.6$~meV) or Co ($T_C=1394$~K,
$E_{ex}=32.3$~meV).

These results suggest that the MR at room temperature could be
strongly improved by using materials with stronger magnetic coupling
$E_{ex}$ or by tweaking the concentration of defects inducing the
magnetic coupling. Also, the exponential decay of the MR with
temperature could be addressed by acceptor doping of the non-magnetic
regions and thereby creating magnetic/non-magnetic p/n junctions.
Although acceptor doping is challenging in ZnO, other semiconductor
materials, e.g. GaN, exhibit defect induced magnetism~\cite{Xu.2014}
and could be used to this end.

\section{Conclusion}
\label{sec:conclusions}

More than ten years after the realization of defect induced magnetism,
we present in this work its practical application in a spintronic
device, which makes use of a spin filtering effect at the interface
between defect-induced magnetic and non-magnetic regions at the
surface of a ZnO microwire. This device exhibits giant
magnetoresistance, which scales with the number of interfaces. The
experimental method described in this work is simple and economical
and can be applied to a large number of materials showing
defect-induced magnetism.

\section{Methods}
\label{sec:experimental}

Li-doped ZnO microwires of $100\,\mu\text{m}$ to $300\,\mu\text{m}$
length and $0.5\,\mu\text{m}$ to $2\,\mu\text{m}$ diameter were
prepared by a carbothermal process~\cite{Lorite.2015,Lorite.2014}. The
Li concentration was chosen following reports in
Ref.~[\onlinecite{Chawla.2009}]. Single wires were selected using an
optical microscope and fixed on a $\text{Si/Si}_3\text{N}_4$
substrate. A dose of $10^{17}\,\text{cm}^{-2}$ of $\text{H}^+$ was
implanted in the extremeties of the wires where gold contacts were
made by e-beam lithography and DC sputtering. The $\text{H}^+$
implantation was realized using a remote hydrogen DC-plasma
chamber~\cite{Lorite.2014} with an implantation energy of
$300\,\text{eV}$ and a current of $60\,\mu \text{A}$. Assuming a
displacement energy of $18.5\,\text{eV}$ (Zn) and $40\,\text{eV}$ (O),
``Stopping and range of ions in matter'' (SRIM)~\cite{Ziegler.2008}
simulations indicate that $\text{H}^+$ is implanted in the top
$10\,\text{nm}$ of the wire surface and that Zn-vacancies
($\text{V}_{\text{Zn}}$) and O-vacancies ($\text{V}_{\text{O}}$) are
created during the implantation. Previous studies indicate that after
implanting a dose of $10^{17}\,\text{cm}^{-2}$, the resistivity of
the Li-doped ZnO material decreases drastically in the 10~nm
subsurface region~\cite{Lorite.2015c}, allowing us the preparation of
ohmic contacts for transport measurements.

In a second step, a PMMA mask consisting of small exposed regions was
created on top of the wires between the contacts using e-beam
lithography (see fig.~\ref{fig:1}~(a)). A dose of
$10^{17}\,\text{cm}^{-2}$ of $\text{H}^+$ was then implanted in the
exposed regions. Due to the creation of $\text{V}_{\text{Zn}}$, the
exposed regions become magnetically ordered~\cite{Lorite.2015}.

In a final step, a dose of $10^{15}\,\text{cm}^{-2}$ of $\text{H}^+$
was implanted in the whole wires. This dose is not sufficient to
induce magnetic order in the regions between the magnetic strips but
reduces the resistivity to enable us to measure a current.

Different wires were prepared using different numbers of strips. Also,
different lengths (parallel to wire axis) of strips were tested,
showing comparable results. In the following, we describe samples from
a batch produced with different numbers of strips of $400\,\text{nm}$
length and $4\,\mu\text{m}$ separation. For example, sample
\textit{S1} was prepared with 37 strips and sample \textit{S2} with 17
strips.

Magnetotransport measurements were performed in a He-cryostat with a
maximum magnetic field of $7\,\text{T}$ at temperatures ranging
between $5\,\text{K}$ and $300\,\text{K}$. Resistance was measured
using a Keithley 6517a Electrometer and IV characteristics were
acquired using a Keithley 220 Current Source and a Keithley 182
Voltmeter.

\section*{Acknowledgement}

This work was supported by the Collaborative Research Center (SFB762)
``Functionality of Oxide Interfaces'' (DFG). Y.K. was partially
supported by the Department of Science and Technology, India
[DST/INSPIRE/04/2015/002938]. Discussions with W. Hergert, W. A.
Adeagbo and A. Ernst are gratefully acknowledged. Patent pending:
German Patent Office application no. 10 2017 001 492.2 (2017).

\vspace{2cm}
\begin{center}
\bf{Supplementary Information} \end{center}

\section{Photoluminescence characterisation}

\begin{figure}[ht]
  \centering
  \includegraphics[width=0.9\textwidth]{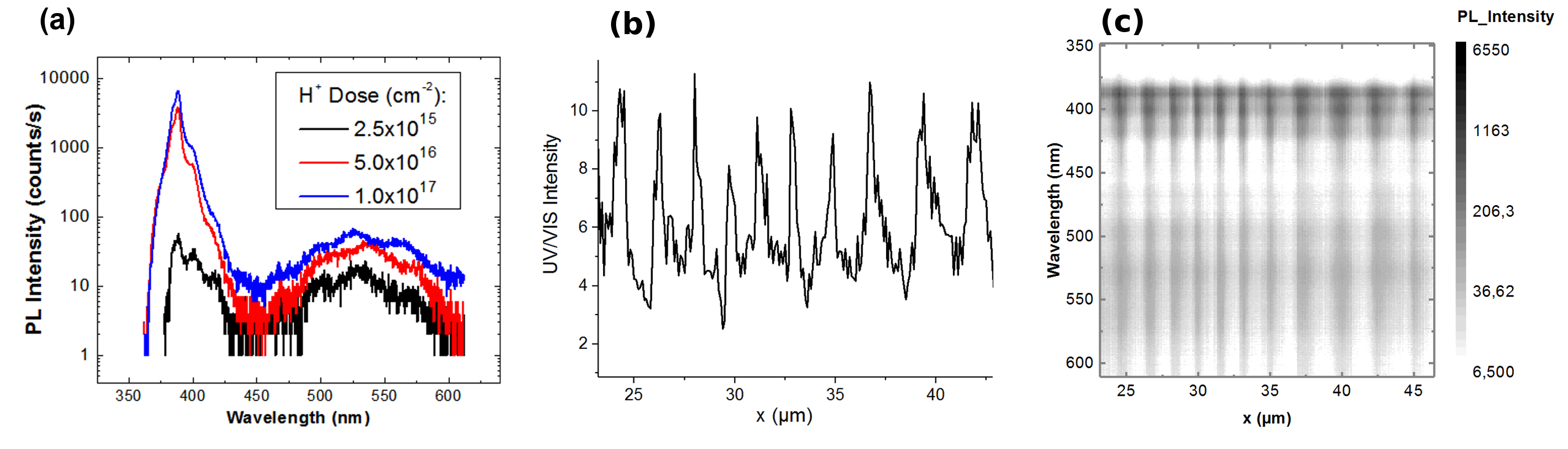}
    \caption{\label{fig:6} \textbf{Photoluminescence characterisation.}
    (a) Photoluminescence spectra of ZnO:Li wires implanted with doses
    of $2.5\times 10^{15}\,\text{cm}^{-2}$,
    $5.0\times 10^{16}\,\text{cm}^{-2}$ and
    $1.0\times 10^{17}\,\text{cm}^{-2}$ of protons respectively. The
    ratio between the intensities of two main emission bands increases
    with the implantation dose. (b) Line scan along the wire S1
    showing the ratio UV/VIS of the intensities of the emissions at
    385nm and 525nm. The regions with high UV/VIS ratio were implanted
    with a dose of $1.0\times 10^{17}\,\text{cm}^{-2}$ of protons, the
    regions with low UV/VIS ratio with a dose of
    $2.5\times 10^{15}\,\text{cm}^{-2}$. (c) Photoluminescence line
    scan along the wire S1, showing the emission spectra in the range
    between 360nm and 610nm.}
\end{figure}

To characterize the magnetic/non-magnetic regions at the surface
of ZnO:Li microwires, we performed micro-photoluminescence
measurements. For reference, we first measured the emission
spectra of a wire after an implantation of $2.5\times
10^{15}\,\text{cm}^{-2}$, $5.0\times 10^{16}\,\text{cm}^{-2}$ and
$1.0\times 10^{17}\,\text{cm}^{-2}$ of protons respectively, as
shown in Fig.~\ref{fig:1}~(a). The wire mainly emits luminescence
in two spectral ranges, namely the UV emission at $\sim
380\,\text{nm}$ attributed to the recombination of free excitons
and their phonon replica~\cite{Lozada.2004, Lu.2013} and several
bands of green luminescence at $450-575\,\text{nm}$ attributed to
intra-bandgap defect levels, such as Zn
vacancies~\cite{Fabbri.2014b}. The ratio UV/VIS between the
emission intensities of the two spectral ranges significantly
increases with the proton implantation dose. We used this fact to
characterize our samples and the magnetic/non-magnetic regions
produced along the wires by proton implantation.
Fig.~\ref{fig:6}~(b) shows the UV/VIS-ratio along the wire S1. The
400nm long magnetic regions implanted with a dose of $1.0\times
10^{17}\,\text{cm}^{-2}$ show high contrast between the UV and
green emission, whereas the $\sim 2\mu\text{m}$ long non-magnetic
regions implanted with a dose of $2.5\times
10^{15}\,\text{cm}^{-2}$ show much less contrast.

\section{Effect of applied bias voltage}

\begin{figure}[h!]
  \centering
  \includegraphics[width=0.9\textwidth]{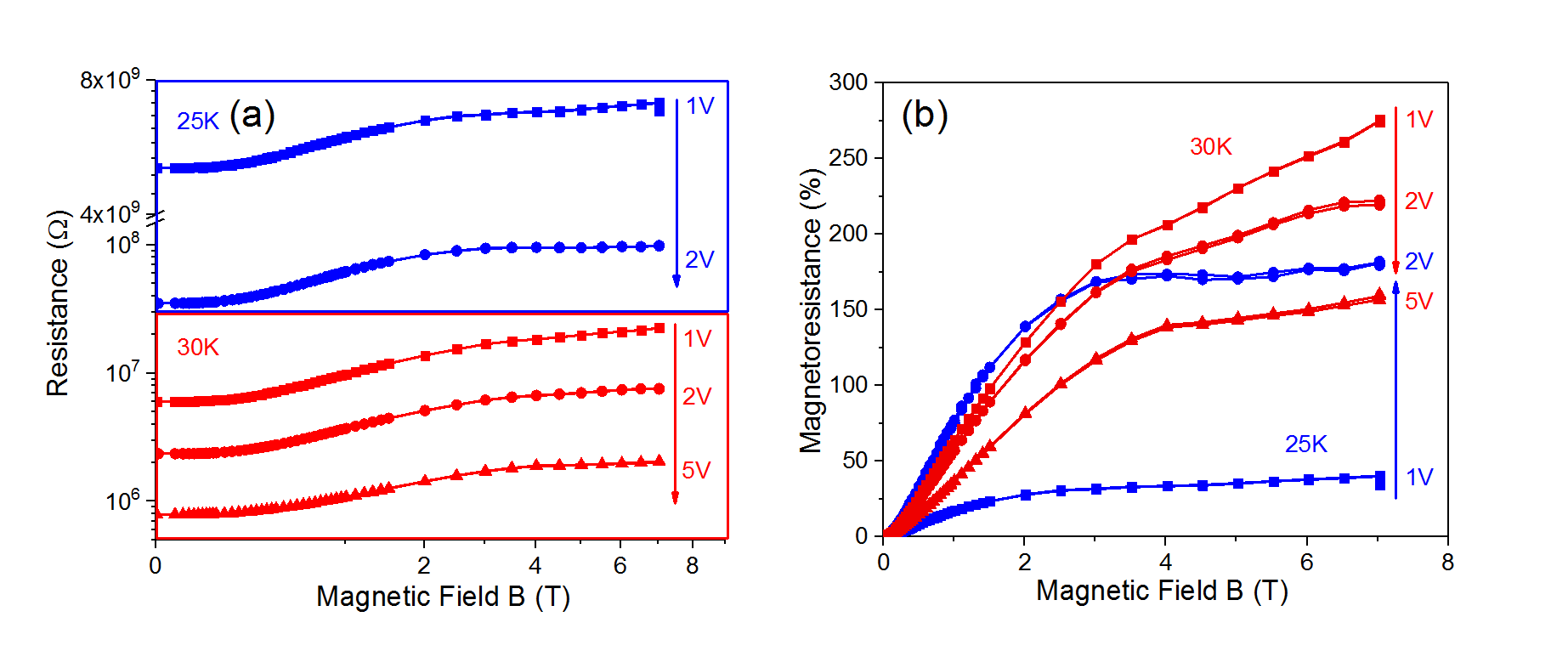}
  \caption{\label{fig:7} \textbf{Bias voltage effect.} (a) Resistance
    of sample S1 as a function of applied magnetic field B measured at
    25K (blue) and 30K (red), with applied bias voltages of 1V, 2V
    and 5V. (b) Magnetoresistance of sample S1 as a function of
    applied magnetic field B measured at 25K (blue) and 30K (red),
    with applied applied bias voltages of 1V, 2V and 5V.}
\end{figure}

In accordance with the IV characteristics shown in the main text,
increasing the applied bias voltage decreases the total resistance
of the wires at all temperatures (see Fig.~\ref{fig:7}~(a)). On
the magnetoresistance however, increasing the applied bias voltage
shows different effects depending on the temperature, as shown in
Fig.~\ref{fig:7}~(b). Below the point of maximal MR, around 30K,
increasing the bias voltage increases the MR while above 30K, the
MR decreases with increasing bias voltage. This behaviour
indicates that the occupation of the spin-split bands changes at
that temperature. Below 30K, the thermal energy is not sufficient
for electrons to reach both spin-split conduction bands. At the
critical point, the lower spin-band is occupied accounting for the
measured current, but the upper spin-band is not and therefore no
spin filtering takes place at the magnetic/non-magnetic junctions.
The application of a bias voltage facilitates the filling of the
upper spin band and enhances the spin filter effect. Above 30K,
both spin bands are filled and the applied bias voltage reduces
the spin filter effect.


%

\end{document}